\documentclass[a4paper,11pt]{article}
\usepackage{pos}
\usepackage{mathrsfs}

\newcommand{\ttr}{\ensuremath{t_\mathrm{tr}}}
\newcommand{\tact}{\ensuremath{t_\mathrm{act}}}

\title{Estimation of the exposure of the TUS space-based cosmic ray observatory}
 \ShortTitle{Estimation of the exposure of the TUS space-based cosmic ray observatory}

 \author*[a,b]{Francesco Fenu}
 \author[c]{Kenji Shinozaki}
 \author[d]{Mikhail Zotov}
 \author[a]{Mario Bertaina}
 \author[b]{Antonella Castellina}
 \author[b]{Alberto Cellino}
 \author[d]{Pavel Klimov}

\affiliation[a]{Università degli studi di Torino,\\
  Via Pietro Giuria 1 10125, Torino, Italy}

\affiliation[b]{INAF, Osservatorio Astrofisico di Torino,\\
  Via Osservatorio 20, 10025, Torino, Italy}

\affiliation[c]{National Centre for Nuclear Research, Cosmic Ray Laboratory,\\
  Lodz, Poland}

\affiliation[d]{Skobeltsyn Institute of Nuclear Physics,\\
  Lomonosov Moscow State University, Russia}

\forColl{JEM-EUSO}

\emailAdd{francesco.fenu@gmail.com}

\abstract{%
	The TUS observatory was the first orbital detector
	aimed at the
	detection of ultra-high energy cosmic rays (UHECRs).
	It was launched on April 28, 2016, from the Vostochny cosmodrome in
	Russia and operated until December 2017. It collected $\sim80,000$
	events with a time resolution of 0.8~$\mu$s.  A fundamental parameter
	to be determined for cosmic ray studies is the
	exposure of an experiment.  This parameter is important to estimate
	the average expected event rate as a function of energy and to
	calculate the absolute flux in case of event detection.  Here we
	present results of a study aimed to calculate the exposure that TUS
	accumulated during its mission.  The role of clouds, detector dead
	time, artificial sources, storms, lightning discharges, airglow and
	moon phases is studied in detail.  An exposure estimate with its geographical distribution is
	presented.  We report on the applied technique and on the
	perspectives of this study in view of the future missions of
	the JEM-EUSO program.}

\FullConference{37$^{\rm{th}}$ International Cosmic Ray Conference (ICRC 2021)\\
		July 12th -- 23rd, 2021\\
		Online -- Berlin, Germany}

\begin{document}
\maketitle

\section{Introduction}

The TUS observatory was launched as a part of the Lomonosov satellite on
April 28$^{th}$, 2016, from the Vostochny cosmodrome in Russia.
The main aim of the mission was to test the observation principle of
a space-based fluorescence
cosmic ray detector.  The satellite flew on a sun-synchronous orbit with
an inclination of 97.3$^\circ$, a 94 minutes period and an altitude of
about 485~km.  The detector could acquire data until December 2017 and
used to cover all latitudes from $-82.7$ to 82.7$^\circ$.  The detector
consisted  of a Fresnel mirror focusing the light onto a
camera. The mirror had an area of $\sim2~\text{m}^{2}$ and the focal
length of 1.5~m.  The field of view was of $9^\circ\times9^\circ$ which
corresponded to an area of approximately
$80~\text{km}\times80~\text{km}$ on ground.  The camera was implemented
as a square matrix of $16\times16$ Hamamatsu R1463 photomultipliers
(PMTs).  The PMTs had a 13~mm diameter multi-alcali cathodes covered by
a UV glass filter and a reflective light guide with a square entrance
with a 15~mm side.  Each PMT covered an area of
$\sim5~\text{km}\times5~\text{km}$ on ground.  The quantum efficiency in
the band 300--400~nm was around 20\%.  The gain of each single channel
was measured before the flight and was found to vary from $5\cdot10^{5}$
to $1.5\cdot10^{6}$~\cite{SSR2017}.  A protection mechanism was
implemented and the gains could be automatically reduced  when the
luminosity was too high.

The readout electronics could operate in four modes specifically
designed for different classes of events.  The Extensive Air Shower
(EAS) mode, which will be relevant for this contribution, was aimed at
the detection of ultra-high energy cosmic rays.  The time sampling in
this case was 0.8~$\mu$s in order to give a good time resolution of
extensive air showers.

The trigger scheme was structured in two steps to allow background
rejection and the acceptance of the cosmic ray events.  A fast ADC
converted analogue signals of PMTs into digital codes with the
resolution of 0.8~$\mu$s.  The digitized signals were summed up on
a sliding window of 16 frames for each photomultiplier.  The integrated
values were compared then with a preset threshold on a moving matrix of
$3\times3$ contiguous pixels.  The first level trigger was activated in
case the threshold was overcome for any of such pixels. The persistency
of such a signal excess was then tested each 16 frames. Once the
persistency was longer than a predetermined value, the second level
trigger was issued.  At this moment the data transfer was initiated.
The \textit{Block of Information} unit, that managed the data acquisition for all scientific devices on board the Lomonosov satellite \cite{SSR2017}, could accept data from TUS at most once in 50-60 seconds.
This external constraint imposed a lower limit to the acquisition dead time of the TUS detector. 

\section{The TUS geometrical aperture}

\begin{figure}[!ht]
	\centering
	\includegraphics[height=6.5cm]{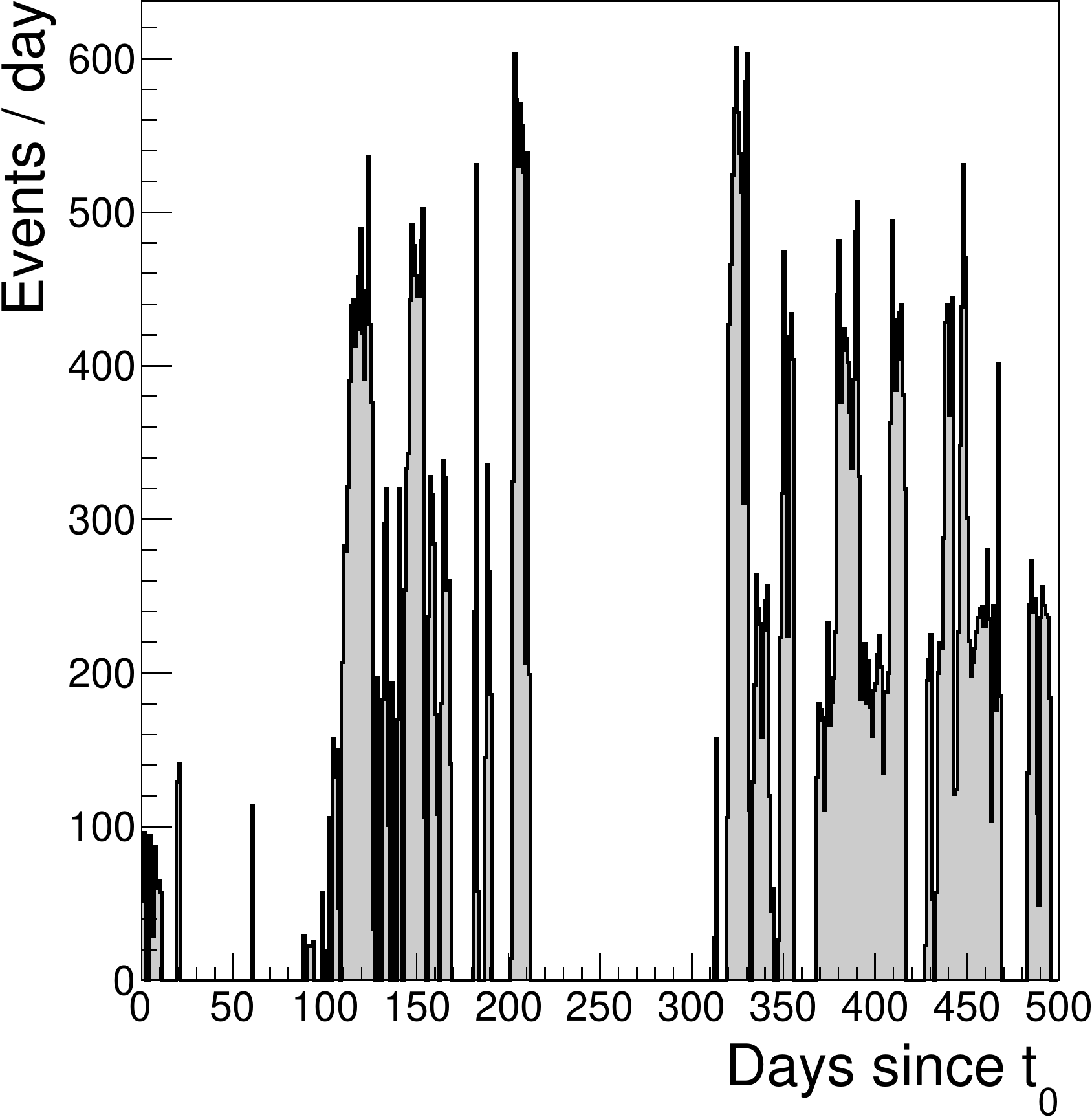}\qquad
	\includegraphics[height=6.5cm]{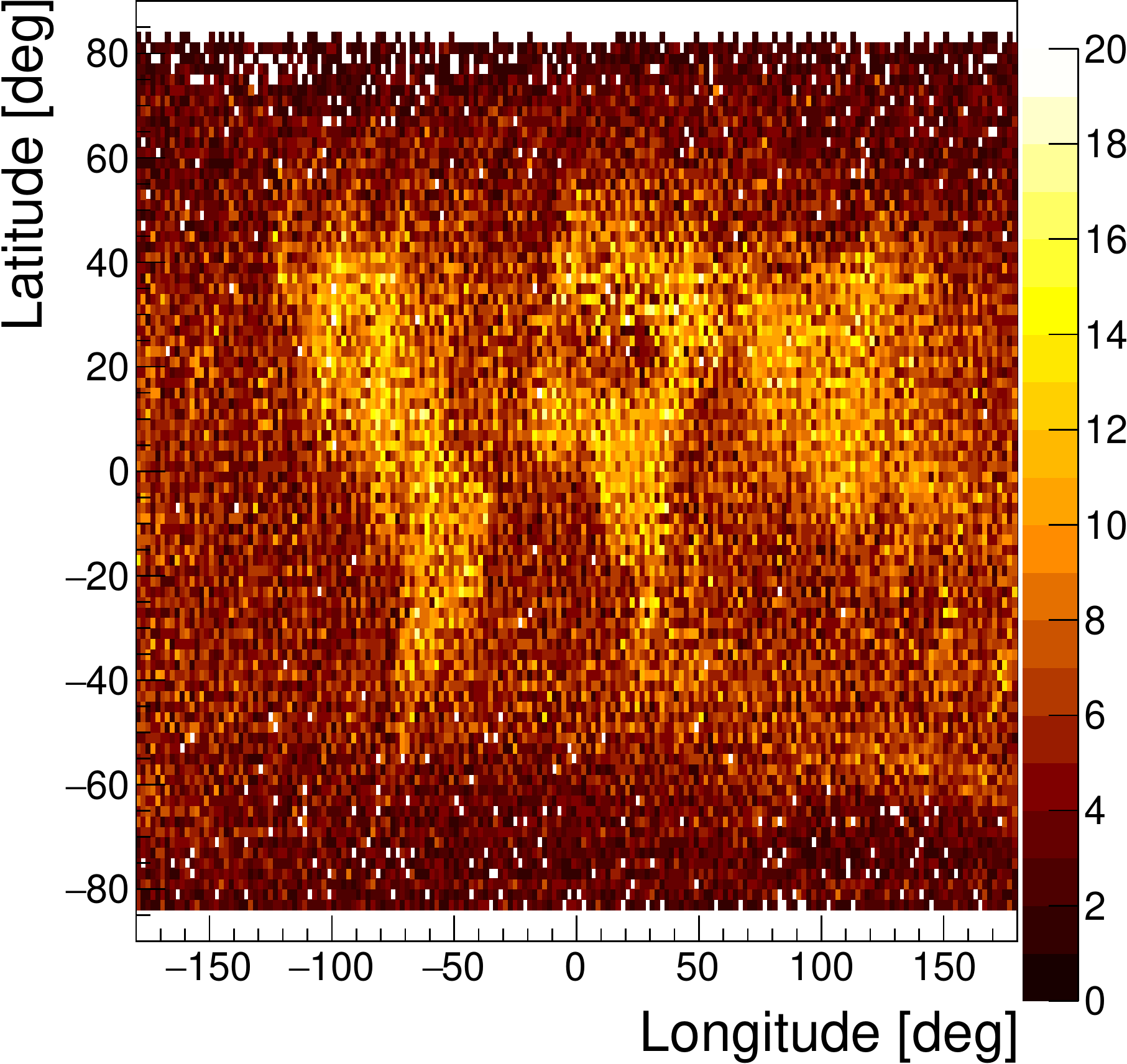}

	\caption{Distribution of the triggered events. Left panel: the time
	distribution from the first day of acquisition. Right panel: the
	geographical distribution of the triggers.}

	\label{fig:DataDistribution}

\end{figure}

The current work is based on events triggered in the EAS
mode in the night segments of the Lomonosov orbits.  
For each trigger, the signal for
all pixels of the focal surface and 256 time frames is available.
Further information like the satellite position and speed vector were
transmitted by the satellite operator Roscosmos.  The distribution of
the triggers in time is shown in the left panel of
Fig.~\ref{fig:DataDistribution}.  In this plot the amount of triggers
per day is shown as a function of the mission time.  The time is
calculated as the number of days since the start of acquisition,
with~$t_{0}$ being May 19, 2016.  Data have been acquired in several
discontinuous sessions, with the highest exposure gathered in Autumn
2016 and in the second half of 2017.  The interruptions are mainly related to the operation in other acquisition modes. The right panel shows the
geographical distribution of the triggers.  As it is clearly apparent,
the triggers were distributed quite uniformly with a higher
concentration over continents.  A notable exception to this is
represented by Antarctica, the arctic and Sahara which remain quiet areas
with the trigger densities comparable to these above the oceans.

\begin{figure}[!ht]
	\centering
	\includegraphics[height=5.5cm]{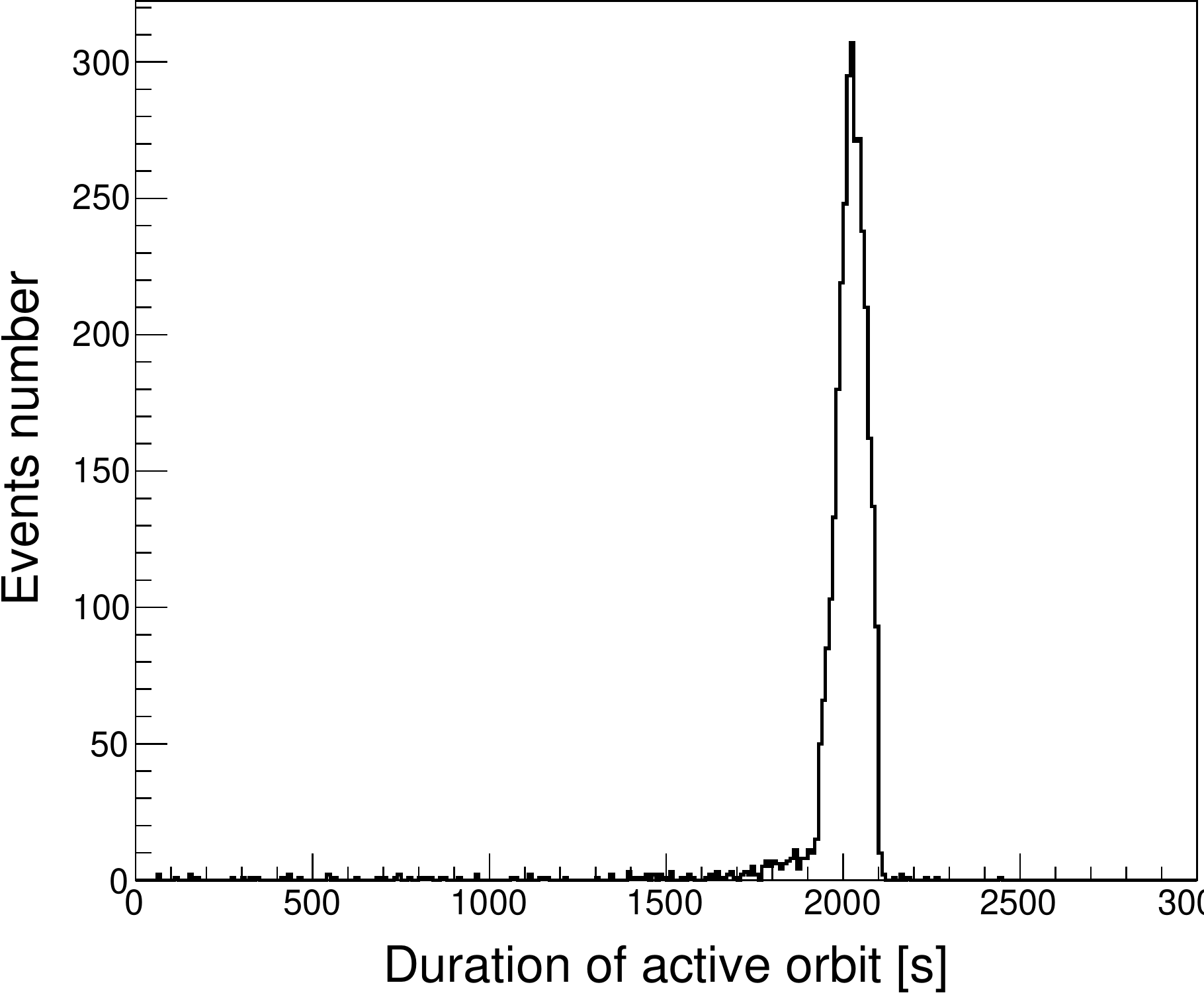}\qquad
	\includegraphics[height=5.5cm]{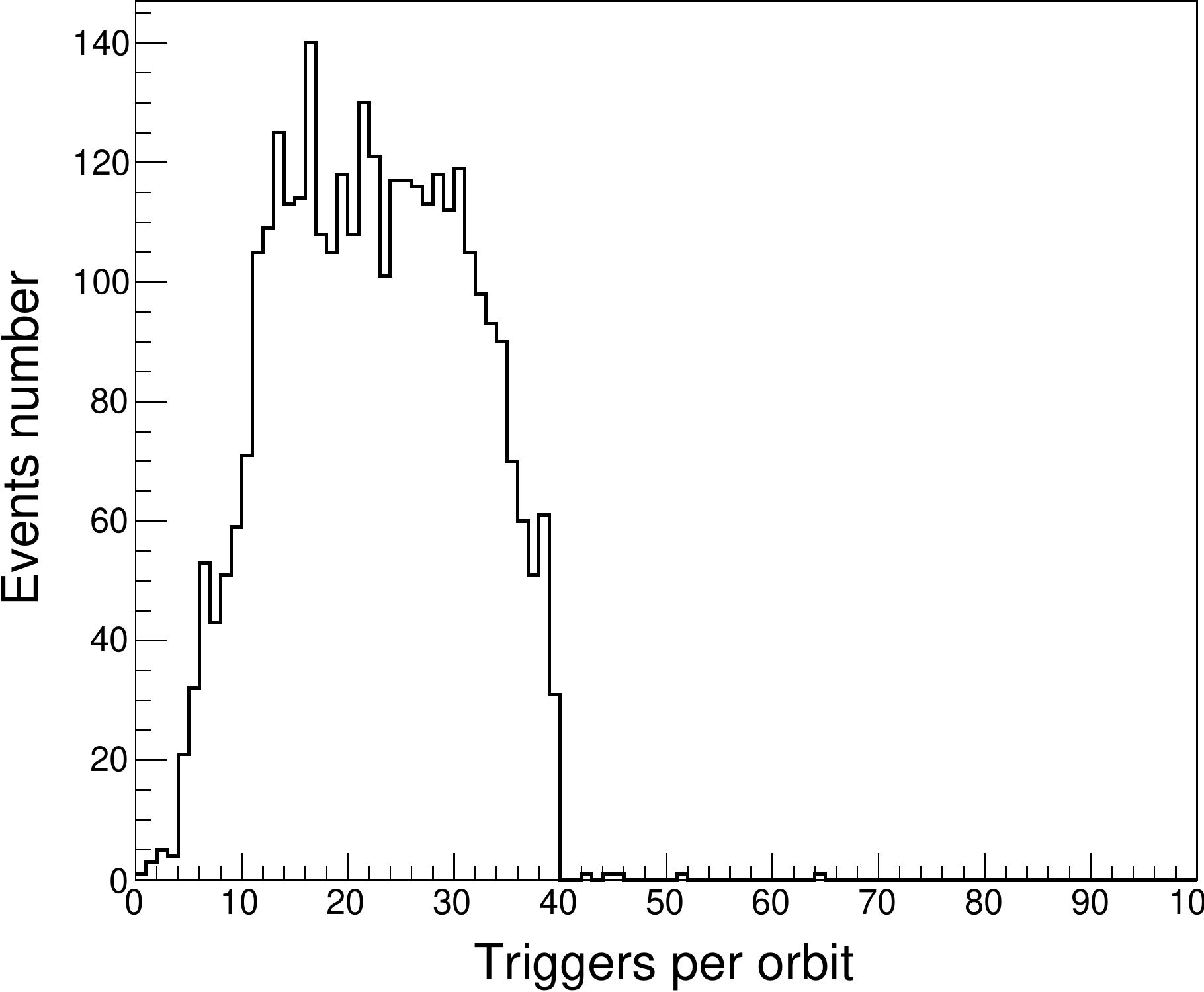}

	\caption{Left side: duration of the night time (or active time) of
	acquisition orbits. Right side: number of triggers per orbit.}

	\label{fig:OrbitDuration}

\end{figure}

Most of the triggers can be grouped in sequences of about 2000~s (33 minutes)
from the first to the last (see Fig.~\ref{fig:OrbitDuration}, left).  As
a matter of fact, 2000~s is the time that the satellite takes to cross
the night side of the Earth and therefore such sequences can be
associated to a single orbit of the satellite.  The number of triggers
per orbit is shown in the right panel of Fig.~\ref{fig:OrbitDuration}.

As we have already mentioned above, TUS had a dead time from 52 to 60
seconds after each trigger, depending on the mission period, and therefore
no orbits with more than $\sim40$ triggers could be generally observed.
An estimate of the active time can be therefore given for each orbit,
under the assumption that the detector has always been in acquisition.
Orbits with a high number of triggers have a very high fraction of dead
time while, on the other hand, orbits with few triggers have a higher
active time.  This way, 3118 orbits with a total acquisition time of 73 full 
days were identified.  A total active time of 31 days is obtained as
soon as the dead time is taken into account. This amounts to $\sim42\%$
of the total acquisition time.
Such an estimate is based only on the identified orbits and could be potentially an underestimate of the real acquisition time.

\begin{figure}[!ht]
	\centering
	\includegraphics[height=12cm]{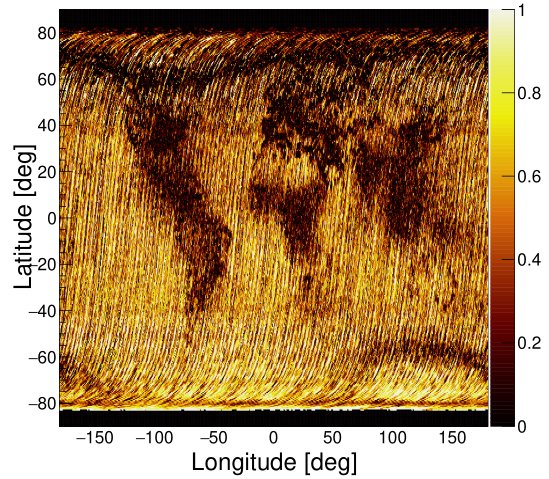}

	\caption{Ratio of active time over the total amount of transit
	time as a function of geographical location.}

	\label{fig:DutyCycle_Dist}
\end{figure}

Thanks to the knowledge of the satellite trajectory, it was possible to
estimate with a $\sim1$ second resolution the status of the
detector for each position on the Earth map.  The geographical
distribution of active time fraction $\psi (\varphi,\lambda) = \tact
(\varphi,\lambda) / \ttr (\varphi,\lambda) $ is shown in
Fig.~\ref{fig:DutyCycle_Dist}.  In this formula, $\tact$ is the amount
of active time, $\ttr$ is the total time integrated in a specific
location of the Earth, $\varphi$ is the latitude and~$\lambda$ is the
longitude.
It can be clearly seen that the presence of a higher trigger rate implies
a higher dead time.  As a consequence of that, populated areas or stormy
regions are basically not contributing to the cumulated exposure.
Aurora ovals are also clearly visible as non-active areas in the polar
regions.  On the other hand, oceans are very quiet areas, where cosmic
ray studies would be favoured.

\begin{figure}[!ht]
	\centering
	\includegraphics[height=5cm]{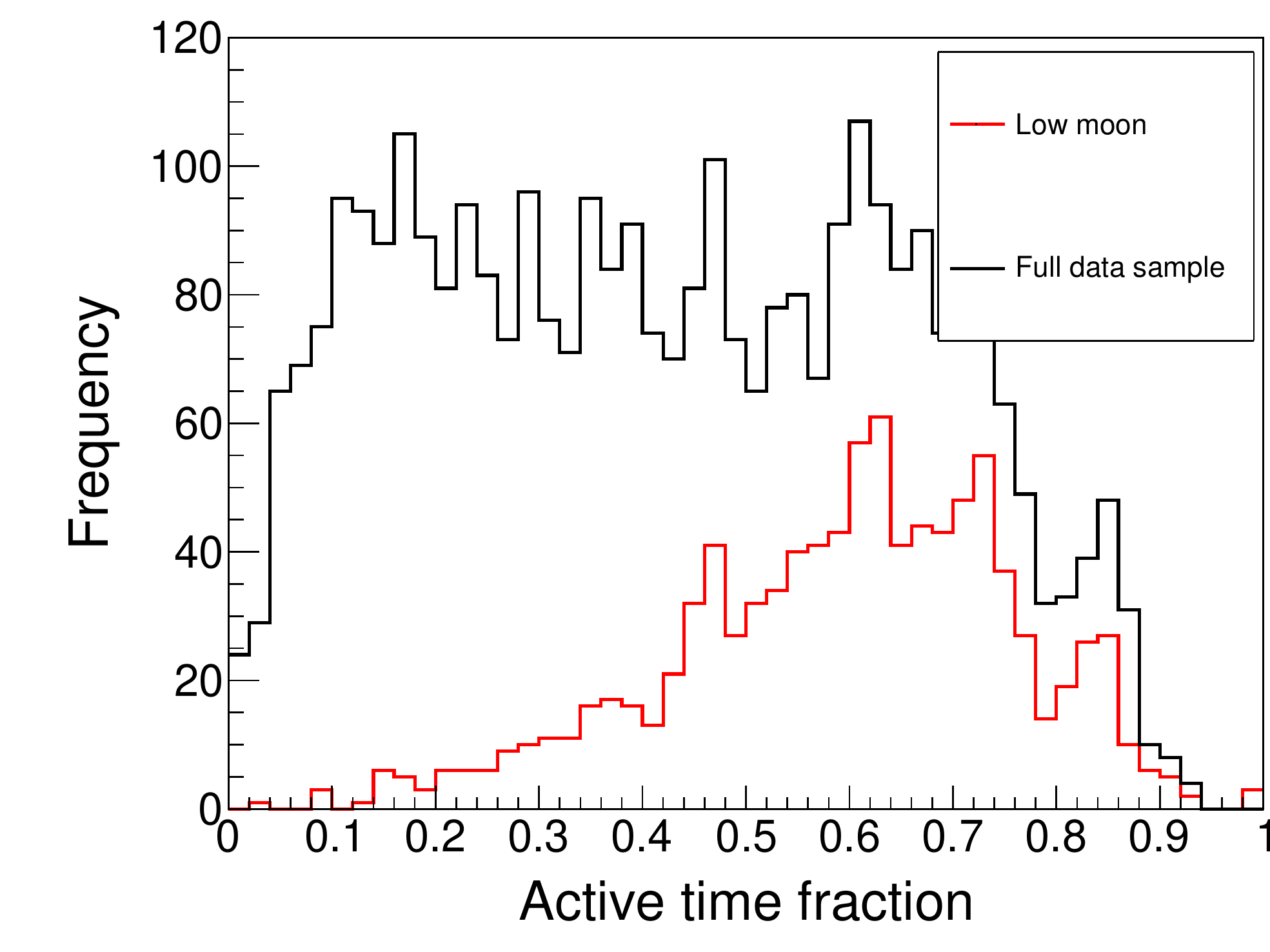}\quad
	\includegraphics[height=5cm]{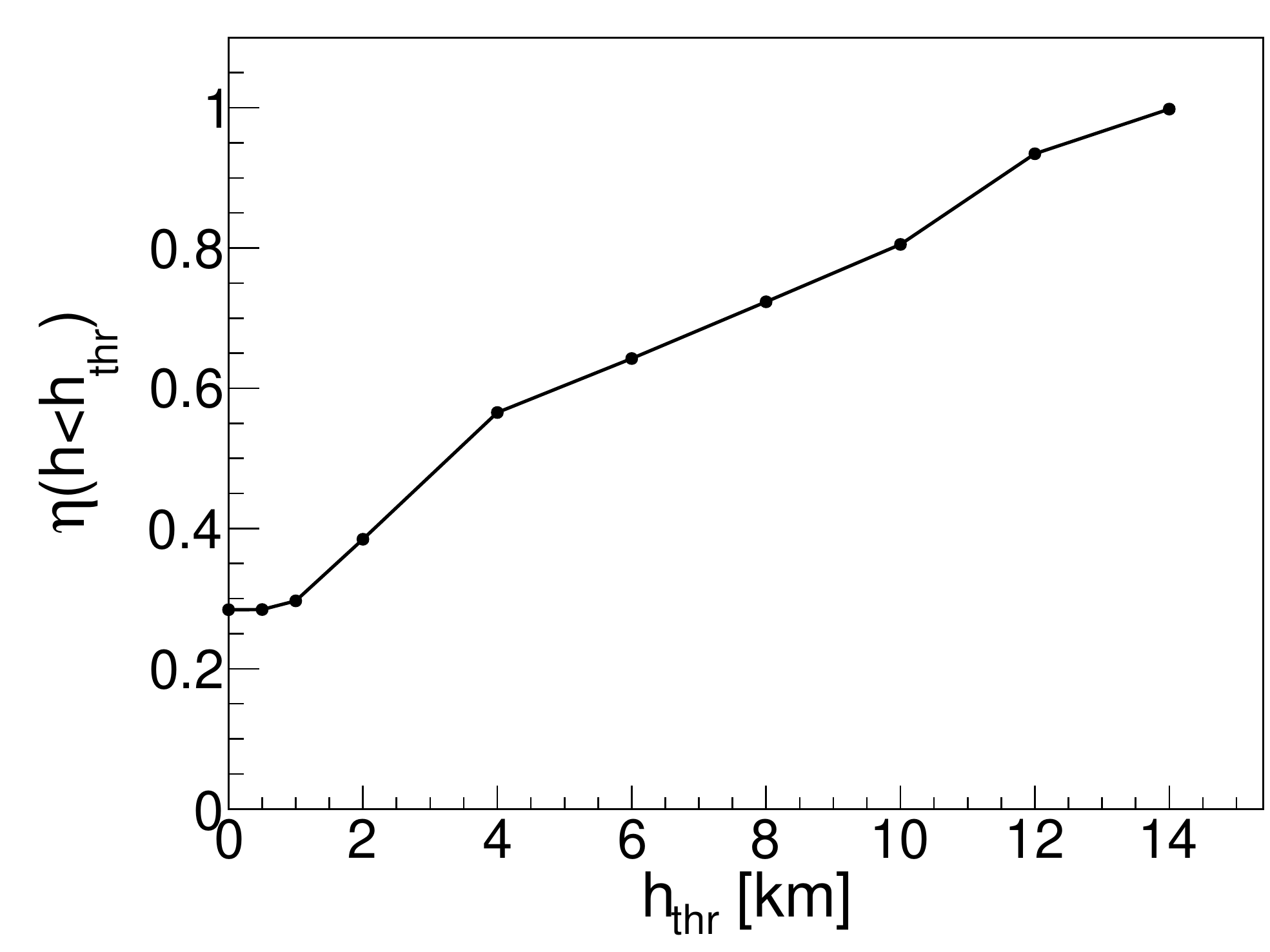}

	\caption{Left panel: active time distribution for each acquisition
	orbit.  The full data set is shown in black.  The distribution with
	low moon illumination is shown in red.  Right panel: fraction of
	triggers with cloud top height under $h_\mathrm{thr}$.}

	\label{fig:Moon}

\end{figure}

Positions of the Sun and the Moon were calculated based on data from the
Japanese Coast Guard \cite{cite:JapanCoastGuard1,
cite:JapanCoastGuard2}.  All triggers could be therefore classified
depending on the Moon illumination.  The left panel of
Figure~\ref{fig:Moon} shows the distribution of the active time fraction
$\psi(\varphi,\lambda)$.  The fraction of data collected when the Moon
was under horizon or under 20\% phase (close to new moon) is shown in
red.  The whole data sample is shown in black.  The presence of low 
or no-Moon illumination is verified in 21.2 full days of acquisition.  The
amount of active time in this condition amounts to 12.9 days, 60\% of
moonless acquisition time.  It is clearly apparent that a cut on the Moon illumination, while decreasing the absolute amount of active time, would increase the data quality.

The cloud condition for each trigger has been estimated based on MERRA data \cite{cite:Merra}.
In Fig. \ref{fig:Moon}  (right side) we plot the fraction of events where the cloud top height is lower than a threshold ($h_{\mathrm{thr}}$).
It can be seen how most ($\sim70\%$) of the triggers are in cloudy condition.
It is therefore crucial to estimate the efficiency for cosmic ray detection in presence of clouds. Simulations will be used for this purpose and we will follow the approach described in \cite{Astrop_exposure}.

\begin{figure}[!ht]
	\centering
	\includegraphics[height=6cm]{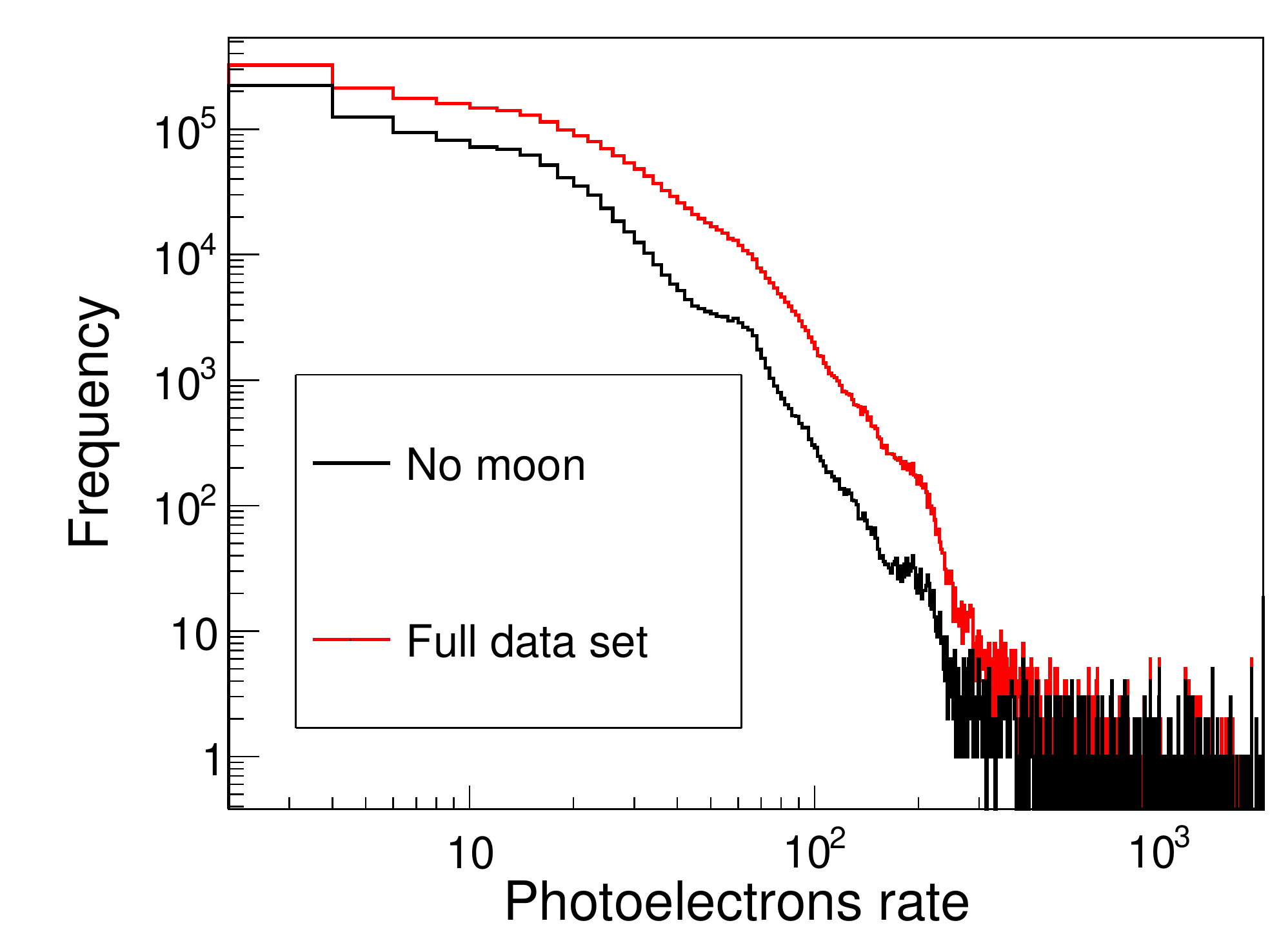}\qquad
	\includegraphics[height=5.8cm]{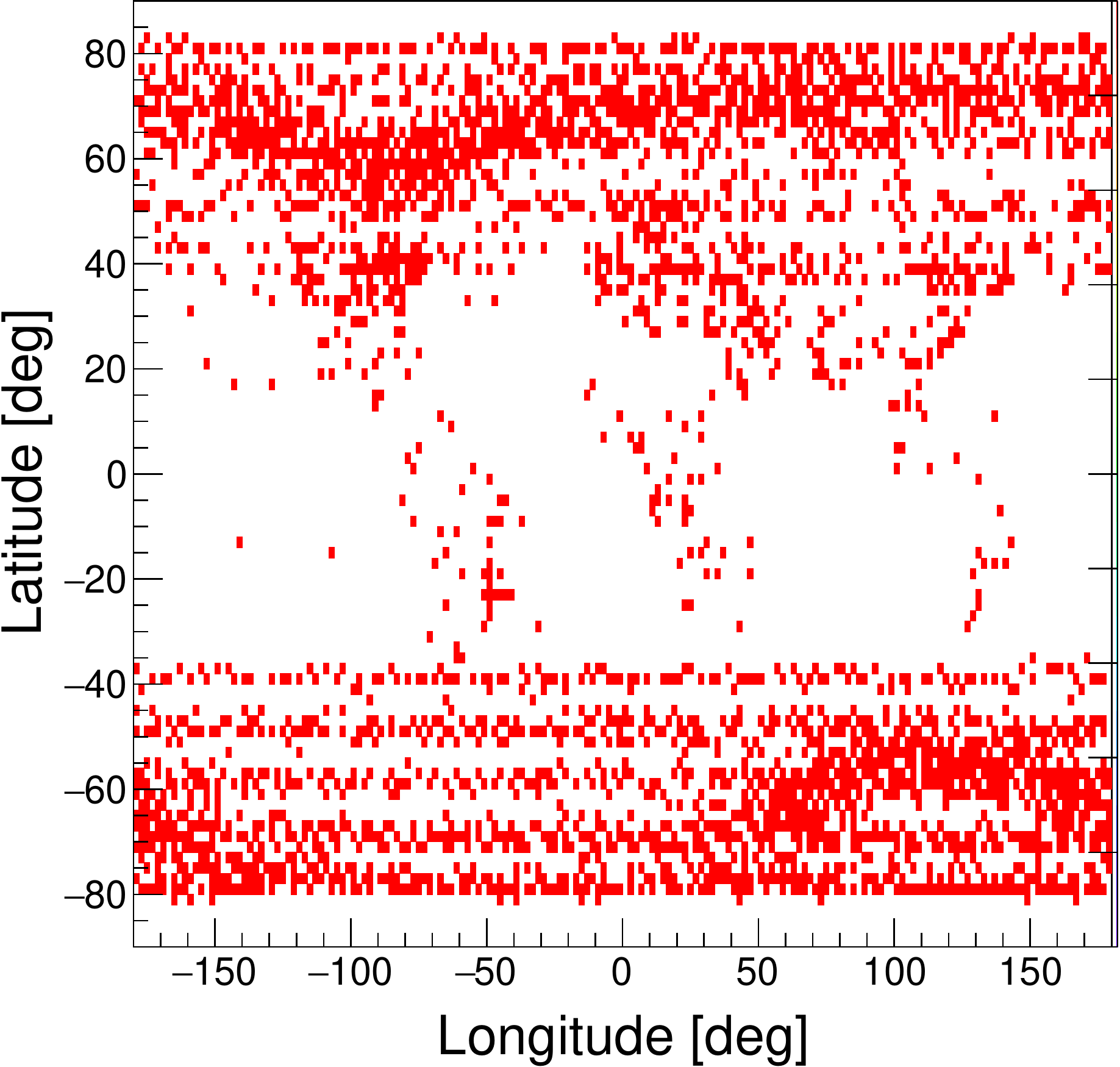}

	\caption{Left panel: estimated average photoelectron rate per pixel
	and time frame from airglow. Right panel: distribution of high
	background rate frames ($N_\mathrm{ph} > 80$).}

	\label{fig:backgroundMeas}

\end{figure}

The signal recorded for each triggered event can be used to estimate the
rate of photoelectrons generated by the airglow emission.  Such information
is fundamental for the estimation of the performances of future
space-based detectors since all trigger algorithms must cope with this
emission.  An accurate estimate of the airglow photon radiance requires
detailed optics simulations which go beyond the scope of this
publication and therefore we will only estimate the rate of
photoelectrons per frame.  Such information will be used in the
simulations to estimate the energy dependence of the exposure.  The
conversion rate from ADC channels to photoelectrons is given by the
following formula:

\begin{equation} \label{eqn:conversion}
	n_{0} = \alpha \frac{q_{e}}{C} \frac{RC}{\Delta t} G,
\end{equation}
where $G$ is the gain of the channel, $C$ and $R$ are the capacitance and
resistance in the anode RC chain of each PMT, $\Delta t$ is the time frame
of TUS and $q_{e}$ is the charge of the electron.  The calculation is
performed only for the cases where the high voltage status flag was set
to 255 ADC channels, or in other terms, where the high  voltage setting
was maximum.  In such cases the measured gain values are reliable and
Eq.~(\ref{eqn:conversion}) can be therefore used.  The 
distribution of the background illumination obtained with the TUS data
is shown in Figure~\ref{fig:backgroundMeas}.

As it can be seen, the rate of the background illumination varies
from~1 to over 100 photoelectrons per frame with some less common cases
up to several thousand photoelectrons.  The right panel of
Figure~\ref{fig:backgroundMeas} shows the geographical location of
events where a rate higher than 80 photoelectrons per frame was
detected.  It is apparent that such extremely high rates were found in
the Aurora ovals or above very densely populated areas.  Several linear
structures can be observed in the Southern hemisphere.  They appeared
to be due to the transition from nocturnal segments of orbits to regions where
the satellite started to receive light from the Sun.  Such data are
collected during the Southern hemisphere summer in a sequence of
acquisition sessions separated by few weeks.  The shift from a dark
region to a bright one causes the generation of triggers in a very
limited space as it can be seen in the figure.  The detector is then
switched to low gain mode  once the satellite enters in the full day
light and all the following triggers are not taken into account anymore.
All such high rate conditions can be identified as extreme cases and do
not reduce the exposure in a significant way.

\section{The TUS trigger performance}

An estimate of the trigger performance is obtained through Monte Carlo simulations. 
Two thousand extensive air showers were injected in an area $A_\mathrm{simu}$ larger than the field of view ($\pm150$~km) to avoid border effects. Showers were simulated with zenith angles from $0^\circ$ to $90^\circ$ and the azimuth from 0 to 360$^\circ$.  The TUS trigger logic was implemented in the ESAF simulation
software~\cite{berat} and used for this estimation.  Several trigger thresholds used in the mission
were tested with an airglow rate of $\sim 18$ photoelectrons per frame.
The estimate of the trigger performance depends on a number of factors, among them the sensitivity of the photodetector, the level of the background illumination
and software parameters of the trigger.  
During an incident described in \cite{JCAP2020}, 20\% of the PMTs were destroyed and sensitivities of the remaining PMTs changed in comparison with pre-flight measurements. A number of attempts of in-flight calibration
have been performed but none of them is fully reliable yet. This
introduces a large factor of uncertainty in estimates of the trigger
threshold that can be wrong by a large amount. 
As a result of our preliminary estimation, we obtain a trigger threshold $\gtrsim400$~EeV. 
 Moreover, the TUS trigger algorithm is more efficient for horizontal showers leading to a higher
fraction of high zenith angle events.  
The majority of the events could indeed trigger only above $40^\circ\text{--}50^\circ$.
This is a consequence of the persistency condition of the trigger that rejects all events lasting for a short time. 

A set of simulations has been performed to estimate the efficiency of
the trigger in cloudy conditions.  Thousand  extensive air showers at fixed energy have been simulated for each cloud top height condition the same way as for clear sky.
Table~\ref{tabFraction} presents the fraction of triggers with clouds
below a specific altitude (as shown in the right panel of
Figure~\ref{fig:Moon} and indicated as $\eta (h<h_\mathrm{thr})$).  The
second row of Table~\ref{tabFraction} shows the ratio of the
efficiency~$\epsilon_{\mathrm{cloud}}$ obtained in cloudy conditions to
the one obtained in clear sky ($\epsilon_{\mathrm{CS}}$).

\begin{table}[!ht]
  \begin{center}
    \begin{tabular}{| c | c | c | c |c | c | c |c | c | c |}
      \hline
                                                        & Clear sky & 1 km    & 2 km   & 4 km   & 6 km   & 8 km  & 10 km   & 12 km  & 14 km \\
      \hline 
       $\eta (h<h_\mathrm{thr} )$                             & 28$\%$    & 29$\%$  & 38$\%$ & 56$\%$ & 64$\%$ & 72$\%$& 80$\%$  & 93$\%$ &  99.8$\%$       \\
      \hline
       $\epsilon_{\mathrm{cloud}}$/ $\epsilon_{\mathrm{CS}}$  &  100$\%$   & 100$\%$  &  83$\%$&53$\%$  & 40$\%$  & 16$\%$& 6$\%$  & 6$\%$ &    0$\%$     \\
      \hline
    \end{tabular}
  \end{center}

  \caption{Reduction of the trigger efficiency due to the presence of clouds with
	respect to clear sky at $2\times 10^{21}$ eV.}
	\label{tabFraction}

\end{table}

A higher cloud top height will cause a significant reduction of the
triggered events given the reduction of the amount of light reaching the
detector.  An estimate of the overall reduction of the exposure in the
whole flight can be given by an average of the trigger efficiency
weighted by the fraction of triggers in each condition. Given the cloud
conditions in the field of view at the time of the flight, an exposure
of 57\% of what is expected for the clear sky case has been estimated.

\begin{table}[!ht]
    \centering
    \begin{tabular}{|c|c|c|c|c|c|}
    \hline
    Airglow rate [ph / frame] & 5 & 18 & 30 & 50 & 100 \\
    \hline
    N$_{\mathrm{Trigg.}}$/ N$_{\mathrm{Trigg.,5}}$ & 100$\%$ & 60$\%$ & 56$\%$ & 18$\%$ & 0$\%$ \\
    \hline
    \end{tabular}
    \caption{Triggers fraction at $2 \times 10^{21}$ eV with respect to the 5 photoelectrons / frame case.}
    \label{tab:my_label}
\end{table}
The efficiency of the trigger was also estimated for different airglow rates. We tested  5, 18, 30, 50 and 100 photoelectrons / frame and estimated the number of triggers in each condition for a fixed energy. The lower rates are characterized by a higher efficiency. By assuming the rate at 5 photoelectrons / frame (N$_{\mathrm{Trigg.,5}}$) to be 100\%, we could estimate the change in the trigger rate shown in Tab. \ref{tab:my_label}.
As expected, we found a strong dependence of the efficiency on the airglow rate. For space-based observatories it will be therefore crucial to estimate the brightness of the field of view and the performances of the trigger in different illumination conditions.

\section{Conclusions}

Thanks to the analysis of the time distributions of the TUS triggers it
was possible to identify a minimum of 3118 complete orbits of data
acquisition.  This has to be considered as a preliminary estimation
given that particularly quiet orbits may not be counted at all in this
analysis.  By considering the number of triggers occurring in such orbits
and the dead time occurring after each trigger, we identified 31 days of
full time acquisition.  The exposure is mainly concentrated on the
oceans, deserts and in Antarctica while populated or stormy areas do not
contribute to the total exposure given the very long dead time of TUS. A
smaller dead time in future detectors will certainly reduce the severity
of this loss of exposure.  Moon-free data are characterized  by a higher
active time fraction.  The estimation of the count rate from airglow
gives a variable rate from a few photoelectrons per frame up to over
100.  The most luminous parts of the Earth are the ones associated to
the Aurora ovals, densely populated areas and to regions
close to the terminator.
The trigger performances must be still studied in detail but preliminary indications point to a trigger threshold $\gtrsim 400$~EeV. 
The presence of clouds reduces the exposure
in this range by $\sim 40\%$.  
As expected, a strong dependence of the trigger performances on the field of view luminosity is found. The design of future missions must take carefully into account the variability of the field of view scenario and the trigger performances for various sky conditions.
Under the above mentioned
assumptions, the purely geometrical exposure of the TUS mission ($\mathscr{E}$), without clouds and for very high event luminosity, can be estimated as $ \mathscr{E} = A_{\mathrm{FOV}}  \times \Omega \times t$.  The geometrical exposure amounts therefore to $\sim 1550~\mathrm{km}^{2}~\text{sr~yr}$.

\section{Acknowledgements}
The Russian group is supported by the State Space Corporation ROSCOSMOS and the Interdisciplinary Scientific and Educational School of Lomonosov Moscow University ``Fundamental and Applied Space Research.''
The Italian group acknowledges financial contribution from the agreement ASI-INAF n.2017-14-H.O.

\clearpage
\section*{Full Authors List: \Coll\ Collaboration}

\begin{sloppypar}
{\small \noindent
G.~Abdellaoui$^{ah}$,
S.~Abe$^{fq}$,
J.H.~Adams Jr.$^{pd}$,
D.~Allard$^{cb}$,
G.~Alonso$^{md}$,
L.~Anchordoqui$^{pe}$,
A.~Anzalone$^{eh,ed}$,
E.~Arnone$^{ek,el}$,
K.~Asano$^{fe}$,
R.~Attallah$^{ac}$
% ...
}
\end{sloppypar}
\vspace*{.3cm}

{ \footnotesize
\noindent
% ...
$^{ac}$ LPR at Dept. of Physics, Faculty of Sciences, University Badji Mokhtar, Annaba, Algeria \\
% ...
$^{ah}$ Telecom Lab., Faculty of Technology, University Abou Bekr Belkaid, Tlemcen, Algeria \\
% ...
$^{cb}$ Universit\'e de Paris, CNRS, AstroParticule et Cosmologie, F-75013 Paris, France\\
% ...
$^{ed}$ Istituto Nazionale di Fisica Nucleare - Sezione di Catania, Italy\\
% ...
$^{eh}$ INAF - Istituto di Astrofisica Spaziale e Fisica Cosmica di Palermo, Italy\\
% ...
$^{ek}$ Istituto Nazionale di Fisica Nucleare - Sezione di Torino, Italy\\
$^{el}$ Dipartimento di Fisica, Universita' di Torino, Italy\\
% ...
$^{fe}$ Institute for Cosmic Ray Research, University of Tokyo, Kashiwa, Japan\\
% ...
$^{fq}$ Nihon University Chiyoda, Tokyo, Japan\\
% ...
$^{md}$ Universidad Polit\'ecnia de madrid (UPM), Madrid, Spain\\
% ...
$^{pd}$ University of Alabama in Huntsville, Huntsville, AL; USA\\
$^{pe}$ Lehman College, City University of New York (CUNY), NY, USA\\
% ...
}

\end{document}